\documentclass[prb,onecolumn,showpacs]{revtex4}
\usepackage{graphicx}

\begin{document}
\title{Anisotropy effects on the Magnetic Excitations
of a Ferromagnetic Monolayer below and above the Curie Temperature}
\author{M. G. Pini$^{1,2}$}
\email{mariagloria.pini@fi.isc.cnr.it}
\author{P. Politi$^{1,2}$}
\author{R. L. Stamps$^{1,3}$}
\affiliation{
$^1$ Istituto dei Sistemi Complessi,
Consiglio Nazionale delle Ricerche,
Sezione di Firenze,
Via Madonna del Piano, I-50019 Sesto Fiorentino, Italy
\\
$^2$ INFM, UdR Firenze, Via G. Sansone 1, I-50019 Sesto
Fiorentino, Italy
\\
$^3$ School of Physics, University of Western Australia,
35 Stirling Highway, Crawley WA 6009, Australia
}
\date{\today}
\begin{abstract}
The field-driven reorientation transition of an anisotropic ferromagnetic monolayer
is studied within the context of a finite-temperature Green's function theory.
The equilibrium state and the field dependence of the magnon energy gap $E_0$
are calculated for static magnetic field $H$ applied in plane along an easy or a hard axis.
In the latter case, the in-plane reorientation of the magnetization is shown to be 
continuous at $T=0$, in agreement with free spin wave theory, and 
discontinuous at finite temperature $T>0$, 
in contrast with the prediction of mean field theory. 
The discontinuity in the orientation angle creates a jump in the
magnon energy gap, and it is the reason why, for $T>0$, the energy does not go
to zero at the reorientation field. 
Above the Curie temperature $T_C$, the magnon energy gap $E_0(H)$ vanishes for $H=0$
both in the easy and in the hard case. As $H$ is increased, the gap is found
to increase almost linearly with $H$, but with different slopes depending on the
field orientation. In particular, the slope is smaller when $H$ is along the hard axis.
Such a magnetic anisotropy of the spin wave energies is shown to persist well
above $T_C$ ($T \approx 1.2~ T_C$).
\end{abstract}
\maketitle

\section{Introduction}
A great experimental achievement is the ability to grow epitaxially
well defined monolayer and sub-monolayer
ferromagnetic films. Such films serve as model systems in which to
study basic aspects of magnetic ordering in low
dimensions as well as to test theories of magnetism. Results
gleaned from these studies have wide impact across several areas from
critical phenomena to electronic band structure
theory. There are also important implications for applications because
the issues involved lie also at the heart of
problems in interface magnetism. A problem of particular importance is
the formation of magnetic anisotropies in thin
transition metal magnetic films.

Critical phenomena in low dimensional magnets and the formation of
magnetic anisotropies are linked together
at a fundamental level. Long range order in two dimensional magnets
depends upon the long wavelength behaviour of
the Goldstone mode of the spin system. This behaviour is strongly
affected by the existence of a zero momentum
energy gap. In most real systems, the largest contribution to this gap
comes from second order corrections to the
exchange energy associated with electronic correlations creating the
ordered state. These corrections
appear in the form of magnetic anisotropies determined by spatial
symmetries of the local atomic environment.

Anisotropy formation is a higher order effect in the sense that the
energies involved are much less than the energy associated with the
ordering temperature $T_C$. In two dimensional systems,
the local fields responsible for magnetic anisotropies are very
different from the local fields existing in three
dimensional bulk systems, and the same reduction in dimensionality has
a profound effect on the critical behaviour
of the system. In real magnetic systems accessible to experimental
study, the formation of magnetic anisotropies and
the critical behaviour of the magnetization are therefore inextricably
linked.

In this paper we show that there are striking features associated with
magnetic anisotropies which appear in spectra
produced by spin wave excitations in two dimensional ferromagnets.
There are very clear and measurable finite temperature effects produced
by anisotropies that can be observed below the ordering temperature,
and we demonstrate that one can also observe easy and hard directionality
in magnetic excitation energies above the ordering temperature.

A few experiments in epitaxial ultrathin ferromagnetic films 
have explored the nature of the phase transition and the persistence
of magnetic anisotropies well above $T_C$;\cite{back:1994,jensen:2003} 
a similar effect was observed in quasi-two-dimensional easy-plane 
ferromagnets.\cite{demokritov} 
The role of the energy gap in stabilizing ferromagnetic order
in two dimensions is well known in critical phenomena, but less well
understood in terms of experiment. Experimentally,
the gap can be observed from the spectra of long wavelength spin waves,
and can be most clearly seen by taking the magnetic system through 
a reorientational transition with the application of a static 
magnetic field along a hard anisotropy direction.\cite{dutcher} 
Theoretically, the equilibrium configuration can be determined, {\it e.g.}   
by the vanishing of the total static torque acting on spins: 
for small fields the magnetization orients between the direction of the field
and the closest symmetry axis of the anisotropy until, 
at some specific critical magnetic field strength $H_c$, 
a reorientation of the magnetization along the field direction occurs.
As regards the spin wave excitations away of the equilibrium state,  
at low temperatures their energy can be easily calculated, 
using the Holstein-Primakoff transformation from spin 
operators to boson operators,\cite{spin-wave-theory}
in the framework of a non interacting spin wave theory.\cite{Yafet,Bruno,noi,Pescia}
For $H=0$ the spin wave energy is essentially determined by 
the anisotropy but, as $H$ increases, the energy barrier to 
deviate away from the easy axis gradually decreases until, 
for $H$ greater than $H_c$, the field itself constitutes an energy barrier. 
This results in a minimum value for the energy of a spin wave at $H_c$.
Within free spin wave theory, valid at low temperatures, 
this minimum is predicted to be exactly zero for zero wavevector.\cite{Pescia}
It is the aim of the present paper to investigate, 
at higher temperatures, the field dependence of the equilibrium  
configuration and of the energy of the magnetic excitations 
of a ferromagnetic monolayer with a uniaxial anisotropy. 
Two different theoretical approaches have been used, namely mean field theory 
(MFT) and Green's function theory (GFT).
The different predictions of these theories are exposed in Sections II and III, 
respectively, both above and below the Curie temperature $T_C$.
Within MFT a constant magnetization modulus is predicted 
for field smaller than the reorientation field, whereas 
within GFT a more accurate treatment of spin fluctuations 
leads to a field dependence of the magnetization modulus 
in the same field range.
As a consequence, for finite temperature $0<T<T_C$, 
we find that the two theories give qualitatively 
different results as regards the order of the field-induced transition:
within MFT the reorientation of the magnetization is continuous 
and the magnon energy gap vanishes at the reorientation field; within GFT the 
reorientation is discontinuous and the gap does not vanish at $H_c$. 
Above $T_C$, both theories are able to account for anisotropy effects 
on the energy of the magnetic excitations, but quantitative differences 
are present. Finally, the conclusions are drawn in Section IV.

\section{Mean field theory}

We consider a magnetic monolayer with Hamiltonian
\begin{equation}
\label{hamiltonian}
{\cal H}= - {J\over 2}\sum_{kl}{\bf S}_k \cdot {\bf S}_l
- K_2 \sum_k(S_k^{{\it Z}})^2-g\mu_B {\bf H}\cdot \sum_k {\bf S}_k
+{1\over 2} {{(g\mu_B)^2}\over {a^3}} \sum_{k\ne l}
{{a^3}\over { r_{kl}^3}}\left\lbrack
{\bf S}_k \cdot {\bf S}_l -
3 {{({\bf S}_k \cdot {\bf r}_{kl})({\bf S}_l \cdot {\bf r}_{kl})}\over
{{\bf r}_{kl}^2}}
\right\rbrack
\end{equation}
\noindent where $J>0$ is the nearest neighbor ferromagnetic exchange
interaction, $K_2>0$ is a uniaxial single-ion anisotropy favoring the
${\it Z}$ direction in the monolayer plane (${\it X}{\it Z}$), and
${\bf H}=(H^{{\it X}},0,H^{{\it Z}})$ is
an external magnetic field  applied in plane: see Fig.~1.
The last term in Eq.~(\ref{hamiltonian}) is the magnetic dipole-dipole interaction,
favoring the alignment of the magnetization within the film plane;
${\bf r}_{kl}$ denotes a vector joining two different lattice sites and
$a$ is the lattice constant. In the following we will assume a simple 
quadratic lattice, so that the number of nearest neighbors is 4.

\subsection{Equilibrium condition}

\begin{figure}
\label{frame}
\includegraphics[width=8cm,angle=0,bbllx=71pt,bblly=177pt,%
bburx=457pt,bbury=562pt,clip=true]{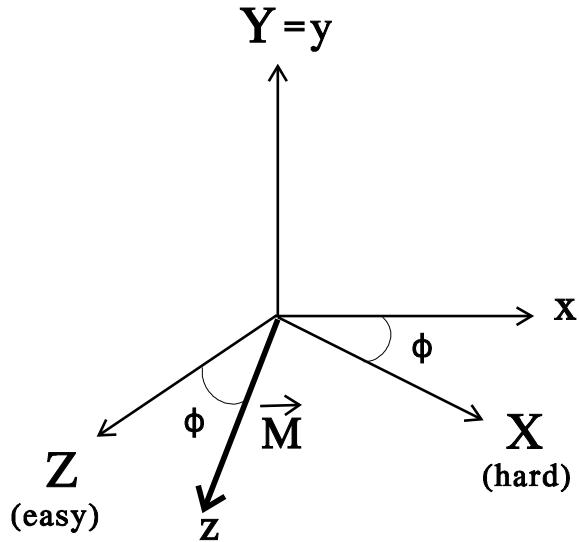}
\caption{Crystallographic (${\it X},{\it Y},{\it Z}$) and local
($x,y,z$) reference
frames used in this paper. The uniaxial in-plane
anisotropy $K_2$ favors the ${\it Z}$ direction within the film plane
(${\it Y}=y=0$)
while a magnetic field is applied in plane along the easy (${\it Z}$)
or the hard (${\it X}$) direction; $\phi$ denotes the angle that the
magnetization {\bf M}, directed along the local axis $z$, forms with
the easy in-plane direction $Z$.}
\end{figure}

The effective field $H_{\hbox{\tiny eff}}$ acting on a spin in the ferromagnetic 
film is composed of an exchange field $H_{ex}={{4J\langle S^z \rangle}\over {g\mu_B}}$; 
a shape demagnetizing field of modulus $4\pi g\mu_B \langle S^z \rangle$ 
favouring the film plane ($Y=0$); a uniaxial anisotropy field 
$H_K={{2K_2\langle S^Z \rangle}\over {g\mu_B}}$
favouring the crystallographic $Z$ direction in the film plane;
an external field $H$ applied within the film plane,
along the easy ($Z$) or the hard ($X$) axis. 
By $\langle S^Z \rangle$ and $\langle S^z \rangle$ 
we denote, respectively, the thermal average of the spin component referred to 
the crystallographic reference frame $X,Y,Z$ and to the local reference frame  $x,y,z$ 
(see Fig. 1). More precisely, $z$ is the equilibrium direction of the 
in-plane magnetization, forming an angle $\phi$ with the easy in-plane 
crystallographic direction ${\it Z}$. In the mean field approximation, 
the thermal average $\langle S^z \rangle$ referred to the local frame can be calculated 
self consistently  in terms of the classical Langevin 
function $\langle S^z\rangle=S {\cal L}(x)$, where ${\cal L}(x)=\coth x-{1\over x}$ and 
$x={{g \mu_B  \mathbf{H}_{\hbox{\tiny eff}}\cdot\mathbf{S^z}}\over {k_B T}}$.

The exchange field is parallel to the magnetization while the shape demagnetizing field
does not provide any preference for a particular in-plane direction.
The equilibrium value for the angle $\phi$
that the magnetization forms with the crystallographic easy axis Z is determined by requiring
the vanishing of the static torque ${\mathbf M} \times ({\mathbf H} + H_K \hat{\mathbf Z})$.

For field along the easy axis, ${\mathbf H}=H^Z \hat{\mathbf Z}$, there is no competition between the
applied field and the anisotropy field, so that $\phi=0$ for any field value.
The thermal average of the magnetization is obtained solving $\langle S^z\rangle=S {\cal L}(x)$ 
with $x=x^{easy}={S\over {k_B T}}(g \mu_B H^Z  + 2 K_2 \langle S^z \rangle +4J \langle S^z \rangle )$.

For field along the hard axis, ${\mathbf H}=H^X \hat{\mathbf X}$, the modulus of 
the anisotropy field is $H_K={{2K_2 \langle S^z \rangle \cos \phi}\over {g\mu_B}}$. 
In this case, the condition for vanishing static torque is
\begin{equation}
H^X \cos \phi={{2K_2 \langle S^z \rangle }\over {g \mu_B}}\cos \phi \sin \phi
\label{equilibrium:mft}
\end{equation}
The thermal average is obtained solving the self consistent equation 
$\langle S^z\rangle=S {\cal L}(x)$ with 
$x=x^{hard}={S\over {k_B T}}(g \mu_B H^X \sin \phi + 2 K_2 \langle S^z \rangle \cos^2 \phi 
+4J \langle S^z \rangle )$. 
As $H^X$ is increased from zero to the critical field value 
$H^X_c={{2K_2 \langle S^z \rangle}\over {g\mu_B}}$, it turns out that $\phi$ 
ranges continuously between 0 and $\pi/2$ and the magnetization modulus 
$M=\langle S^z \rangle$ is independent of the field intensity in the 
whole field range $0 \le H^X \le H^X_c$ (see Fig.~2a).

\begin{figure}
\label{FIG_GF_MF}
\includegraphics[width=18cm,angle=0,bbllx=64pt,bblly=327pt,%
bburx=445pt,bbury=598pt,clip=true]{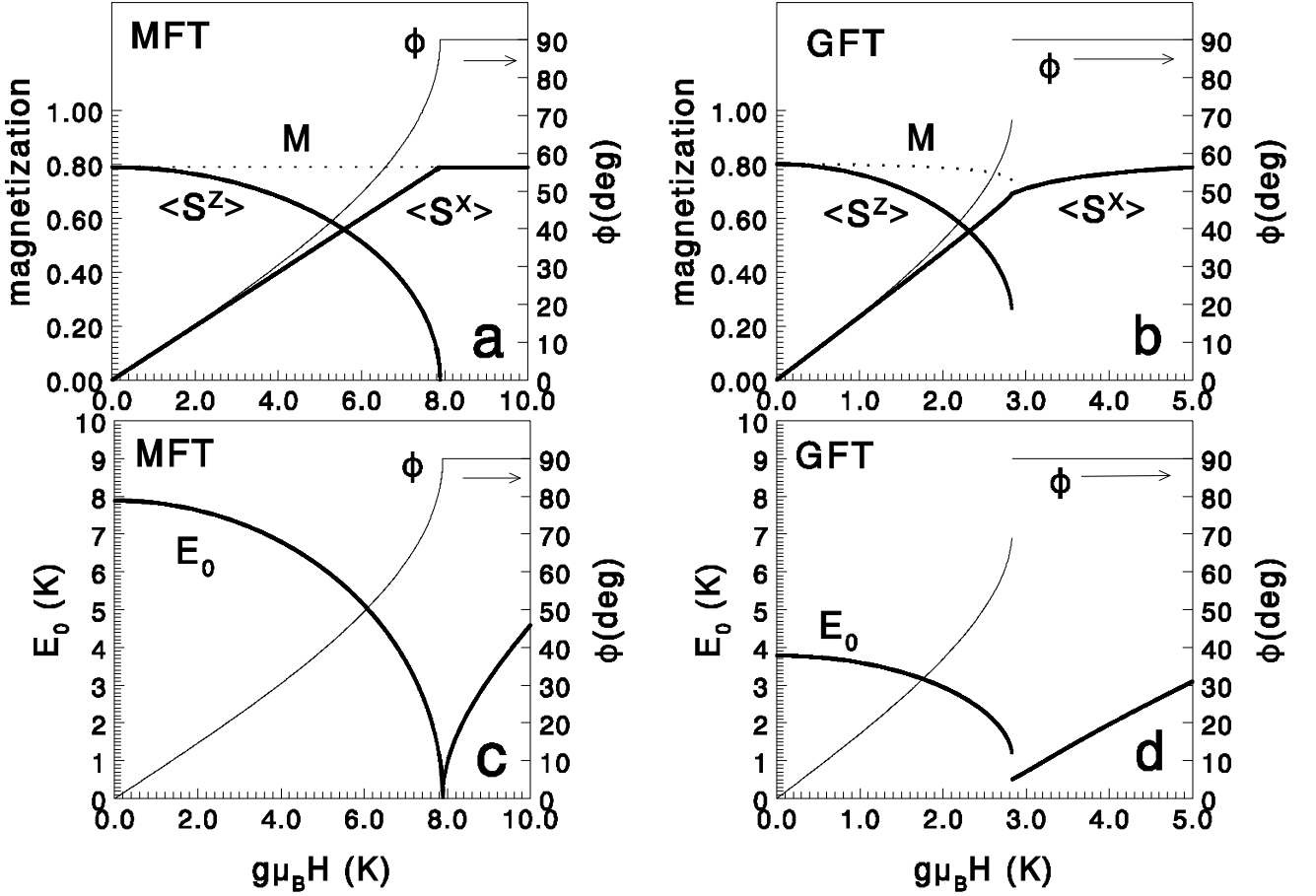}
\caption{Comparison between the results of mean field theory (MFT, left) 
and Green's function theory (GFT, right) for the magnetization components (above)
and the energy gap (below) versus the intensity $H$ of a magnetic field 
applied in plane along the hard direction ($X$), for fixed 
temperature $T/T_C=0.5$.
Figs.~2a,b display the calculated equilibrium angle $\phi$ 
(thin full line, referred to the right vertical axis), 
magnetization modulus $M=\langle S^z \rangle$ (dotted line) and 
magnetization components, $\langle S^Z \rangle=
M\cos\phi$ and $\langle S^X \rangle=M\sin\phi$ (thick full lines), versus $H$. 
Figs.~2c,d display the calculated equilibrium angle $\phi$ (thin full line) 
and energy gap $E_0$ (thick full line) versus $H$.
The Hamiltonian parameters used for the calculations are
$J=500$ K, $K_2=5$ K, $4\pi w f_w=1$ K, $S=1$, providing $T_C=670$ K in 
mean field theory and $T_C=502.4$ K in Green's function theory.
}
\end{figure}

The latter feature can be readily accounted for observing that, 
upon substituting the equilibrium condition appropriate for $0 \le H^X \le H^X_c$ (namely
$\phi=\sin^{-1}\left( {{g \mu_B H^X}\over {2 K_2 \langle S^z \rangle}} \right)$)
in the expression for $x^{hard}$, one obtains that the magnetization modulus 
$M=\langle S^z \rangle=S {\cal L}(x)$ is determined by the solution of the 
self consistent equation 
$x={{S^2}\over {k_B T}} (2K_2+4J) (\coth x-{1\over x})$,  
(where the field intensity $H^X$ does not explicitly appear)
in the field range from $H^X=0$ up to $H^X=H^X_c$. In contrast, 
for $H^X>H^X_c$ the magnetization is field dependent since 
one has $\phi=\pi/2$ and $x={{S}\over {k_B T}}[g\mu_B H^X+4JS(\coth x-{1\over x})]$.
Notice that the field dependence of $M$ for $H^X>H^X_c$ is not apparent 
from Fig.~2a since it is very weak for the chosen parameter values.

For zero field, one has $x={S\over {k_B T}}(4J+2 K_2) \langle S^z \rangle$ and 
the Curie temperature $T_C$ can be calculated, as usual, by requiring 
that the initial slope of ${\cal L}(x)$ is equal to 1, 
{\it i.e.} to the slope of $\langle S^z \rangle/S$. One obtains
\begin{equation}
k_B T_C={1\over 3} S^2 (4J+2K_2).
\end{equation}

\subsection{Resonance frequency}

The resonance frequency can now be calculated by solving 
the linearized Bloch torque equations 
$d\mathbf{M}/dt=-\gamma \mathbf{M}\times\mathbf{H}_{\hbox{\tiny eff}}$
where $\gamma$ is the gyromagnetic ratio.
For field along the easy axis, ${\mathbf H}=H^Z \hat{\mathbf Z}$,
the time varying magnetization can be assumed to be of the form 
$\mathbf{M}=\left(\hat{\mathbf{X}}m_{X}+\hat{\mathbf{Y}}m_{Y}\right)e^{-i \omega t}
+\hat{\mathbf{Z}} g\mu_B\langle S^z \rangle$, 
while the effective field is 
\begin{equation}
\mathbf{H}_{\hbox{\tiny eff}}= - \hat{\mathbf{Y}}4\pi m_{Y}
+\hat{\mathbf{Z}}\left(
H^Z+ {{2 K_{2}\langle S^z\rangle}\over {g\mu_B}} \right)
\label{eq:easy}.
\end{equation}
We note that the effective exchange field was not included 
in Eq.~(\ref{eq:easy}) since it is parallel to the magnetization, 
thus it does not contribute to the uniform mode frequency.
The resonance frequency is then readily found to be
\begin{equation}
\omega_0^{easy}=\gamma 
\left\lbrack
H^Z+{{2 K_{2}\langle S^z\rangle}\over {g\mu_B}}
\right\rbrack^{1\over 2}
\times 
\left\lbrack
H^Z+{{2 K_{2}\langle S^z\rangle}\over {g\mu_B}}+4\pi g\mu_B\langle S^z \rangle
\right\rbrack^{1\over 2}
\end{equation}
where $\langle S^z \rangle=S {\cal L}(x)$ and 
$x=x^{easy}={S\over {k_B T}}(g \mu_B H^Z  + 2 K_2 \langle S^z \rangle +4J \langle S^z \rangle )$.
It turns out that $\omega^{easy}$ monotonously increases as $H$ is increased 
(not shown in Fig.~2c).

For field along the hard axis, ${\mathbf H}=H^X \hat{\mathbf X}$,
the time varying magnetization is
$\mathbf{M}=\left(\hat{\mathbf{x}}m_{x}+\hat{\mathbf{y}}m_{y}\right)e^{-i \omega t}
+\hat{\mathbf{z}} g\mu_B\langle S^z \rangle$ (where $x$, $y$ and $z$ denote local axes), 
while the effective field is 
\begin{equation}
\mathbf{H}_{\hbox{\tiny eff}}=\hat{\mathbf{X}} H^X - \hat{\mathbf{Y}}4\pi m_{Y}
+\hat{\mathbf{Z}}{{2 K_{2}}\over {g\mu_B}} \left(
\langle S^z \rangle \cos \phi - {{m_x}\over {g\mu_B}} \sin \phi
\right)
\label{eq:hard}
\end{equation}
(where $X$, $Y$, $Z$ denote crystallographic axes).\cite{note1}
Expressing the effective field in local coordinates and solving the 
linearized Bloch torque equations, the resonance frequency is found to 
display a non monotonous behavior (see Fig.~2c), different for $H^X$ below and 
above the reorientation field $H^X_c={{2K_2 \langle S^z \rangle}\over {g\mu_B}}$
\begin{eqnarray}
\omega_0^{hard}&=&\gamma 
\left\lbrack 1+ {{4\pi g^2 \mu_B^2}\over {2K_2}}\right\rbrack^{1\over 2} 
\times \left\lbrack
\left( {{2K_2\langle S^z\rangle}\over {g\mu_B}} \right)^2 -(H^X)^2
\right\rbrack^{1\over 2}
,~~~{\rm for}~~H^X<H^X_c
\cr
\omega_0^{hard}&=&\gamma 
\left\lbrack H^X+4\pi g \mu_B \langle S^z \rangle \right\rbrack^{1\over 2} 
\times \left\lbrack H^X- {{2K_2\langle S^z \rangle}\over {g\mu_B}}\right\rbrack^{1\over 2}
,~~~{\rm for}~~H^X>H^X_c
\end{eqnarray}
where $\langle S^z \rangle=S {\cal L}(x)$, 
$x=x^{hard}={S\over {k_B T}}(g \mu_B H^X \sin \phi + 2 K_2 \langle S^z \rangle \cos^2 \phi 
+4J \langle S^z \rangle )$ and $\phi$ is given by the equilibrium condition,  
Eq.~(\ref{equilibrium:mft}).
It is worth noticing that the finite temperature mean field theory predicts 
a vanishing resonance frequency of the uniform mode just at $H^X_c$ as 
a consequence of the continuous in-plane reorientation of the magnetization.

\subsection{High temperature expansions}

The behavior of spin wave energies in the high temperature regime above
$T_C$ is made particularly interesting by the reduced dimensionality of the system.
In two dimensional magnets, the spin system is highly susceptible to
perturbations and can respond strongly even at
temperatures above $T_C$. This means that a two dimensional
ferromagnet in the paramagnetic regime can still produce
large scale correlations when in a static applied field. Furthermore,
these correlations are sensitive to the orientation of the applied field 
relative to the symmetry axis of the anisotropy.

For temperatures above $T_C$ a finite magnetization is
possible only in the presence of an applied field. Furthermore, since
the reorientation field is found to vanish at $T=T_C$, the field-induced
magnetization will align parallel to the applied field.
A simple estimate of the resonance frequencies above $T_C$ can thus be
made expanding the Langevin function for high temperatures (small $x$) 
${\cal L}(x) \approx {x\over 3} - {{x^3}\over {45}}$.
For field along the easy axis (${\mathbf H}=H^Z \hat{\mathbf Z}$)
and along the hard axis (${\mathbf H}=H^X \hat{\mathbf X}$), respectively,
the resulting resonance frequency is
\begin{equation}
\omega_0^{easy}= \gamma H^Z \left\lbrack 1+\left( 4 \pi g \mu_{B}  + 
\frac{2 K_{2}}{g\mu_{B}} \right)  \Gamma_{+}(T) \right\rbrack^{{1\over 2}}
\times \left\lbrack 1+\frac{2 K_{2}}{g\mu_{B}}  \Gamma_{+}(T)\right\rbrack^{{1\over 2}}
\label{omegaeasy},
\end{equation}
\begin{eqnarray}
\omega_0^{hard}=\gamma H^X \left[ 1- {{2 K_{2}}\over {g\mu_{B}}}
\Gamma_{-}(T) \right]^{{1\over 2}}
  \left[ 1+ 4 \pi g\mu_{B} \Gamma_{-}(T)\right]^{{1\over 2}}
\label{omegahigh}.
\end{eqnarray}
\noindent with
$\Gamma_{\pm}(T)=\left(\frac{1}{3} \pm \frac{4}{45}\frac{K_{2}}{k_{B}T}\right)
  {{g\mu_B S^{2}}\over {k_{B}T}}$.
The frequencies have a linear dependence
on field with different temperature dependent slopes. The slope is smaller 
for the field in the hard direction because in that case the anisotropy reduces 
the torque acting on the spins, thereby reducing the susceptibility with 
respect to the easy case.

\section{Green's function theory}

Although mean field theory has proved useful to study 
reorientation transitions in multilayer films,\cite{usadel}
it is well known that it grossly underestimates the effect 
of single site anisotropy in Heisenberg models\cite{devlin} 
and that it leads to estimates of the critical temperature
generally too high due to the neglect of spin correlations.
As the reorientation field depends on anisotropy and on 
the temperature dependence of magnetization, one can expect 
that mean field theory is not capable of providing results that
can be compared quantitatively with experiments in high quality 
epitaxial systems. 
A Green's function theory using the random phase 
approximation\cite{zubarev,callen} 
appears to be a better choice since it provides a more correct treatment 
of both issues, namely single site anisotropy and spin correlations. 

Recently, a number of Green's function theories were proposed 
by different authors\cite{froebrich,froebrich2,froebrich3,froebrich_annals,nolting} 
to treat the problem of the field-induced transition in ultrathin ferromagnetic films.
A thorough discussion of the limits of the approximations used in 
Ref. \onlinecite{froebrich} to treat the reorientation transition 
was provided by the authors of Ref. \onlinecite{nolting}, who proposed 
a theory based on a generalization of the Callen\cite{callen} decoupling valid for an arbitrary 
direction of the external field.

In this work we present a Green's function approach which was
developed independently from Ref. \onlinecite{nolting}.
We start from the outset for rotation 
of the quantization axis for fields below the reorientation transition: 
{\it i.e.}, we write the Green's functions equations of motion 
in the local reference frame. 
Next, by a careful treatment of the uniaxial anisotropy 
through a generalized Callen decoupling\cite{callen} 
of the higher order Green's functions, we obtain - in a very simple way 
even in the case of field applied along the hard axis - both the equilibrium condition 
and the frequency of the magnetic excitations at finite temperature $T>0$.
As a consequence of the generalized Callen's decoupling, the validity of our
theory is limited to low values of the ratio $r=K_2/J$ between the uniaxial 
anisotropy and the exchange constant. For high values of $r$, it is advisable 
to use a more refined Green's function approach\cite{devlin} where the single ion 
anisotropy terms are treated exactly by introducing higher order Green's functions and, 
taking advantage of relations between products of spin operators, a closed set 
of $2S$ equations of motion is obtained for the anisotropy Green's functions. 
Also the quantum Monte Carlo approach in Ref. \onlinecite{henelius} 
is expected to be appropriate for large values of the ratio $r$, while,
in the opposite limit of small $r$, finite size effects strongly
influence the `critical region' around the reorientation transition).

Coming back to our Green's function approach, we observe that 
complete agreement with Ref. \onlinecite{nolting} is obtained  
for the equilibrium condition\cite{note_nolting_1}  while  
different expressions are found for the energies of the magnetic 
excitations.\cite{note_nolting_2} 
Apparently, this discrepancy is due to the fact that in the Green's function 
theory of Ref. \onlinecite{nolting} some extra approximation was made, 
consisting in the treatment of the anisotropy in terms of an effective field, 
as far as the magnetization and the magnon energies are concerned.

In order to study the reorientation transition and the
field dependence of the magnon energies, we have taken particular care in correctly 
recovering  the results of non interacting spin wave theory
(see Appendix B) in the low temperature limit, when the local thermal averages 
tend to constant values, $\langle S^z \rangle \to S$ and $\langle (S^z)^2\rangle \to S^2$.
This limit is not recovered\cite{note_nolting_3} by the approach 
in Ref.~\onlinecite{nolting}.
In particular, we will show (see later Eq.~\ref{E0hard}) 
that the well-known feature\cite{morrish} of a zero frequency of the 
uniform ($k=0$) magnon mode at the reorientation field, 
predicted also by free spin wave theory (see Appendix B),  
is correctly recovered by our Green's function approach in the $T \to 0$ limit.

\subsection{Equilibrium condition and energy of the magnetic excitations}

In order to calculate the equilibrium condition and the energies 
of magnetic excitations in the monolayer described by Eq.~(\ref{hamiltonian})
we use a Green's function formalism\cite{zubarev} supplemented by a random phase
approximation (RPA) with a careful treatment of the uniaxial anisotropy 
through a generalized Callen decoupling.\cite{callen}
Details of the calculation in the absence of magnetic dipole-dipole 
interactions are given in Appendix A. Hereafter we only quote the final results, 
where dipole-dipole interactions were included in an approximate way 
(see Appendix C for details). 
Taking a rotation procedure of the quantization axis $z$ (see Fig.~1),
we obtain the RPA equilibrium condition in the general case of in-plane field
with crystallographic components $H^X$ and $H^Z$ (see Eq. (\ref{equilibrium}) in Appendix A)
\begin{equation}
\label{eq_main}
 2 K_2 f_S \langle S_i^z \rangle \sin \phi \cos \phi
+g \mu_B (H^{{\it Z}} \sin \phi - H^{{\it X}} \cos \phi)=0
\end{equation}
where the factor $f_S$, 
\begin{equation}
f_S=1-{1\over {2S^2}}\left\lbrack S(S+1)-\langle S_i^z S_i^z\rangle \right\rbrack, 
\end{equation}
is required in order to preserve the kinematic consistency of the magnetic excitations 
(see later).

It is worth noticing that in the $T \to 0$ limit one has $\langle S^z_i \rangle \to S$, so that
the RPA equilibrium condition  reduces to the
equilibrium condition one would obtain in the framework of free spin wave theory,
using the standard Holstein-Primakoff transformation for spin operators,
by imposing the vanishing of the terms linear in $a$ and $a^+$
in the boson Hamiltonian.\cite{spin-wave-theory} The physical meaning of 
this equilibrium condition is the neglect of correlations between 
the longitudinal ($z$) and transverse spin components. In fact, we remind that 
in the framework of free spin wave theory the longitudinal ($z$)
spin component has a boson representation in terms of zero- and two-boson operators
while the transverse ($x,y$ or, equivalently, $\pm$) spin components 
are expressed in terms of one-boson operators (see Appendix B for details).
Similarly, the RPA equilibrium condition was obtained  by imposing 
the vanishing of the Green's function $G^z_{ij}(\omega)$ 
(see Eq. (\ref{Gz}) in Appendix A), which correlates the longitudinal 
and the transverse spin components. 

For field applied along the hard axis ($H^Z=0$ and $H^X \ne 0$),
the equilibrium angle is given by 
$\phi = \sin^{-1}({{g \mu_BH^X}\over {2K_2 f_S \langle S_i^z \rangle}})$ for 
$H^X \le H^X_{c}$ and $\phi = {{\pi}\over 2}$ for $H^X > H^X_{c}$.
The reorientation of the magnetization within the film plane
occurs at a critical value of the applied field
$H^X_{c}={{2K_2 f_S \langle S_i^z \rangle}\over {g\mu_B}}$.
Notice that, as in the mean field theory, the reorientation field is a function of temperature,
but the RPA condition contains the additional temperature dependent correlation function
$\langle S_i^z S_i^z\rangle$ in the kinematic consistency factor $f_S$. 
The latter factor is obtained by a careful decoupling of the
higher order Green's functions coming from the uniaxial anisotropy,\cite{callen}
and is required in order to preserve the kinematic consistency of the magnetic excitations:
{\it i.e.}, for $S=1/2$, the uniaxial anisotropy does not contribute to the magnon energy
(see later, Eqs. (\ref{magnon},\ref{Ak},\ref{Bk})). In fact, for $S=1/2$ one has $f_S=0$;
for $S=1$, which is the simplest non-trivial case, one has $f_S={1\over 2} \langle S_i^z S_i^z \rangle$.

The general form of the magnon energies in the momentum representation is, including 
also the magnetic dipole-dipole interaction (see Appendices A and C for details) 
\begin{equation}
\label{magnon}
E_{{\bf k}_{\Vert}}=\sqrt{A_{{\bf k}_{\Vert}}^2 -B_{{\bf k}_{\Vert}}^2}
\end{equation}
where
\begin{equation}
\label{Ak}
A_{{\bf k}_{\Vert}}=
4J\langle S_i^z \rangle (1-\gamma_{{\bf k}_{\Vert}})+
2K_2 f_S \langle S_i^z\rangle \cos^2 \phi
- K_2 f_S\langle S_i^z \rangle  \sin^2 \phi
+{1\over 2} 4 \pi w f_w \langle S_i^z\rangle
+ g\mu_B (H^{{\it X}}\sin\phi +H^{Z}\cos\phi)
\end{equation}
\begin{equation}
\label{Bk}
B_{{\bf k}_{\Vert}}=K_2 f_S \langle S_i^z \rangle \sin^2 \phi
+ {1\over 2} 4 \pi w f_w \langle S_i^z \rangle
\end{equation}
In Eqs.~(\ref{Ak},\ref{Bk}), $\gamma_{{\bf k}_{\Vert}}={1\over 2}[\cos(k_x a)+\cos(k_z a)]$
is the geometric factor for the simple quadratic (s.q.) lattice and  $w={{(g\mu_B)^2}\over {a^3}}$ 
denotes the strength of the magnetic dipole-dipole interaction, while  
the factor $f_w=1.078$ comes from the calculation of dipolar sums 
for the s.q. lattice.\cite{Yafet,Bruno,noi}
Notice that the wavevector dependence of the dipolar sums was neglected
for the sake of simplicity (see the Appendix C for details).  Such an approximation
for the dipolar interaction is certainly plausible in the presence of a sufficiently
large gap in the spectrum.

\begin{figure}
\label{fig_gap}
\includegraphics[width=18cm,angle=0,bbllx=57pt,bblly=332pt,%
bburx=515pt,bbury=631pt,clip=true]{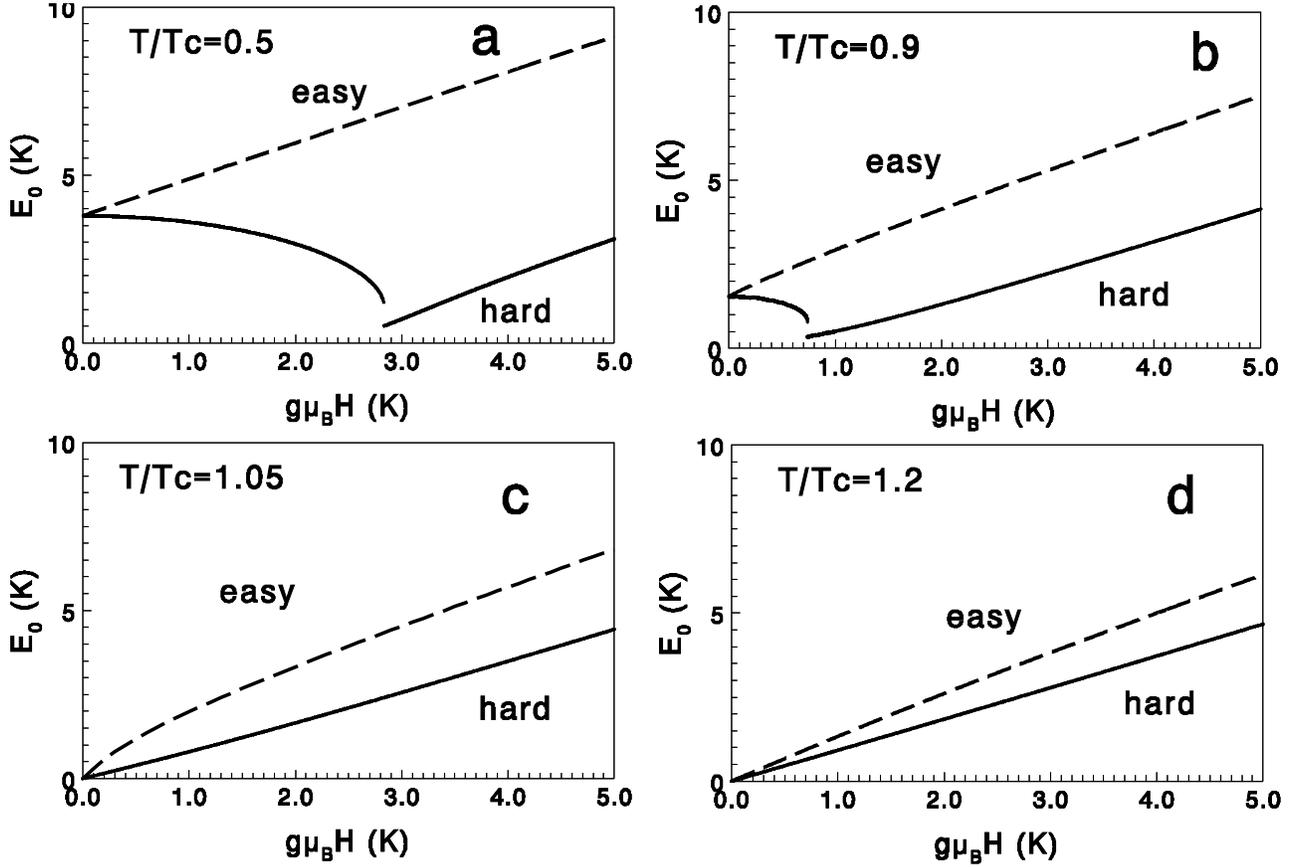}
\caption{Green's function calculation of the energy gap 
for field applied along the hard axis (full lines) 
and the easy axis (dashed lines) at four different temperatures
($T/T_C$=0.5, 0.9, 1.05 and 1.2, respectively).
With the Hamiltonian parameters used for the calculations, 
$J=500$ K, $K_2=5$ K, $4\pi w f_w=1$ K, $S=1$ (giving $T_C=502.4$ K), 
a magnetic anisotropy of the spin wave energies is found to persist
up to $T/T_C=1.2$.}
\end{figure}

For $S=1$ the relevant thermal averages in the RPA theory are
(general expressions for any value of $S$ can be found in Appendix A)
\begin{equation}
\langle S_i^z \rangle= {{1+2 \Phi(T)}\over {1+3\Phi(T)+3\Phi^2(T)}}
\label{Sz}
\end{equation}
\begin{equation}
\langle (S_i^z)^2 \rangle = S(S+1)- (1+2\Phi(T)) \langle S_i^z \rangle 
= {{1+2 \Phi(T)+2\Phi^2(T)}\over {1+3\Phi(T)+3\Phi^2(T)}}
\label{SzSz}
\end{equation}
where the temperature dependence is self-consistently given by 
($N$ is the total number of spins in the s.q. lattice)
\begin{equation}
\label{Phi(T)}
\Phi(T)={1\over {2N}}\sum_{{\bf k}_{\Vert}} \left\lbrack
{{A_{{\bf k}_{\Vert}}}\over {E_{{\bf k}_{\Vert}}}}
\coth \Big({1\over 2} \beta E_{{\bf k}_{\Vert}}\Big) -1
\right\rbrack.
\end{equation}

For zero field, in the $T \to T_C$ limit, one has $\langle S_i^z \rangle \to 0$,
so that $E_{{\bf k}_{\Vert}}$ vanishes with it and $\Phi(T)$ diverges.
Thus from Eqs.~(\ref{Sz}) and (\ref{SzSz}) one has $\langle S_i^z \rangle \to {2\over {3\Phi(T)}}$ and
$\langle S_i^z S_i^z \rangle \to {2\over 3}$, respectively.
By equating the expression (\ref{Phi(T)}) (expanded for $E_{{\bf k}_{\Vert}} \to 0$)
with $\Phi(T)={2\over {3\langle S_i^z \rangle}}$, one obtains the following expression
for the Curie temperature of the $S=1$ ferromagnetic monolayer in zero field
\begin{widetext}
\begin{equation}
k_B T_C= {2\over 3} \left\{
{1\over N} \sum_{{\bf k}_{\Vert}}
{
{4J(1-\gamma_{{\bf k}_{\Vert}}) +{2\over 3}K_2 +{1\over 2} 4 \pi w f_w}\over
{
\Big( 4J(1-\gamma_{{\bf k}_{\Vert}}) +{2\over 3}K_2 +4 \pi w f_w \Big)
\Big( 4J(1-\gamma_{{\bf k}_{\Vert}}) +{2\over 3} K_2 \Big)
}
}
\right\}^{-1}
\end{equation}
\end{widetext}

\subsection{Discussion}

In Fig.~2b we show the self-consistently calculated equilibrium angle $\phi$, 
magnetization  $M=\langle S^z \rangle$, and magnetization components
$\langle S^Z \rangle=\langle S^z \rangle \cos \phi$
and $\langle S^X \rangle=\langle S^z \rangle \sin \phi$, 
as a function of the magnetic field $H$, applied along the hard axis, 
for fixed finite temperature ($T/T_C=0.5$). 
Correspondingly, in Fig.~2d the self-consistently calculated energy gap 
$E_0$ is shown as a function of $H$.

In contrast with the mean field result shown in Fig.~2a, 
in the framework of Green's function theory we find that the orientation angle $\phi$ 
varies discontinuously at the reorientation field and that the magnetization modulus
$M=\langle S^z \rangle$ decreases as $H$ is increased from 0 to
the critical value $H^X_{c}={{2K_2 f_S \langle S_i^z \rangle}\over {g\mu_B}}$.
This is related to the more accurate treatment of spin correlations in 
the Green's function theory with respect to mean field theory. 
In fact we observe that both the temperature and field dependence 
of $\langle S^z \rangle$ occurs through the quantity 
$\Phi(T)$ in Eq. (\ref{Phi(T)}), which is determined by a summation over all magnon modes.
At fixed temperature, the strongest contribution to $\Phi(T)$ 
comes from the $k=0$ mode since this has the lowest energy for 
$H^X \to (H^X_{c})^{-}$. This explains why $M=\langle S^z \rangle$ 
decreases in this limit. In turn, the orientation angle 
$\phi$, whose value is determined by the value of 
$\langle S^z \rangle$, undergoes an abrupt variation at 
the reorientation field. 

As a consequence of the discontinuity in the orientation 
angle $\phi$, the energy gap $E_0$ displays a discontinuous jump 
at the reorientation field $H^{{\it X}}_{c}={{2K_2 f_S \langle S_i^z \rangle}\over {g\mu_B}}$
(see Fig. 2d) and the energy does not go to zero at the reorientation field. 
This is in contrast with the finite 
temperature mean field result displayed in Fig.~2b. 

It is important to note that the softening of the $k=0$ magnon energy 
at the reorientation field\cite{morrish} is correctly recovered by 
our Green's function approach in the $T=0$ limit. In fact for $T \to 0$ and $S=1$ 
one has $\langle S^z \rangle=\langle S^z S^z \rangle \to 1$
and the energy of the $k=0$ magnon mode exactly coincides with  
the result of non interacting spin wave theory for field $H^X$ applied 
along the hard direction
(see Appendix B for details)
\begin{eqnarray}
\label{E0hard}
E_0&=&\left\lbrack (K_2  )^2
 - (g\mu_B H^X)^2 \right\rbrack^{1\over 2}
\times \left\lbrack 1+ {{4\pi w f_w}\over {K_2}}
\right\rbrack^{1\over 2},~{\rm for}~H^X \le H^X_{c}
\cr
E_0&=&\left\lbrack
g\mu_B H^X -K_2  \right\rbrack^{1\over 2} \times
\left\lbrack
g\mu_B H^X +4 \pi w f_w \right\rbrack^{1\over 2},~{\rm for}~H^X > H^X_{c}.
\end{eqnarray}
This energy is immediately seen to vanish continuously
at the reorientation field $H^X_{c}={{K_2 }\over {g \mu_B}}$.

The magnitude of the gap energy at zero field is interesting to consider.
The zero field gap value is a measure of the effective anisotropy energy, 
and is strongly temperature dependent:
the zero field gap value vanishes for $T>T_C$, as well as the reorientation field.
This is clearly a consequence of the Callen decoupling\cite{callen} 
of the higher order Green's functions generated by the uniaxial anisotropy,
yielding a decoupling factor proportional to the magnetization, see 
Eq. (\ref{kinem_anyspin}). Physically this is related to the absence of 
a collectively ordered magnetic state for $T > T_C$. 
However, as discussed by Jensen {\it et al.},\cite{jensen:2003}
the vanishing of the {\it effective} anisotropy for $T>T_C$ 
does not indicate that the {\it microscopic} anisotropy vanishes either:
a single magnetic moment in the paramagnetic state is still 
subject to the anisotropy even if the net magnetization is zero.
While in Ref. \onlinecite{jensen:2003} such a ``paramagnetic anisotropy" 
was revealed by magnetic susceptibility measurements, in the present paper 
we rather focus on the gap energy behavior. 
This feature is illustrated in Fig.~3, where the gap energy 
is plotted as a function of the magnetic field applied 
in the easy and in the hard direction, 
at various fixed temperatures, both below and above $T_C$.
For $T\ge T_C$, the easy- and hard-direction energies are each
zero at zero field, and their field dependence becomes more and more linear
as $T$ increases. However, the calculated slopes are different, 
in agreement with the mean field argument: 
the hard axis alignment slope is smaller than the easy axis alignment slope.
Finally, it is worth noticing that for a realistic value of the ratio 
$K_2/J$ in an ultrathin film ($\approx {1\over 100}$, as in our numerical calculations), 
the magnetic anisotropy of the energy gap is expected to persist well above $T_C$
(up to $T/T_C \approx 1.2$, see Fig.~3d). 

\section{Conclusions}
Using a Green's function theory with random field approximation we 
have shown that the field driven reorientation transition of a 
ferromagnetic monolayer with uniaxial anisotropy 
is expected to be continuous at $T=0$ 
(in agreement with free spin wave theory) and discontinuous at finite temperature.
In contrast, mean field theory predicts a field driven reorientation
transition continuous at any temperature. 
Below the critical temperature $T_C$, the Green's function theory 
predicts a magnetic excitation gap (the resonance mode) which does not completely soften 
at the reorientation field. 

The question of the order of the transition should be considered 
as a still open question, because it is not fully clear if and how 
the decoupling scheme affects that behavior. One step in this 
direction might be accomplished performing a decoupling 
of the Green's functions on a higher level.\cite{junger}
It would also be interesting to analyze the order of this type of transition
in experimental systems ({\it e.g.}, by ferromagnetic resonance 
measurements of the uniform mode as a function of the field).

The magnetization of a two dimensional ferromagnet is
highly susceptible above the critical temperature $T_C$. This
means that a large magnetization can be induced with relatively
small applied fields in the paramagnetic regime.
Such effects were demonstrated in beautiful experiments
on ultrathin Fe/W(110) films by Back et al.\cite{back:1994}
More recently, such ``paramagnetic anisotropy" effects 
were investigated by means of magnetic susceptibility 
measurements on ultrathin Co/Cu(1~1~17) films 
by Jensen {\it et al.}\cite{jensen:2003}
In our case, this means that a two dimensional film
with well defined anisotropies can support magnetic
excitations in the paramagnetic regime with energies that depend
upon the orientation of the applied field relative to the
anisotropy symmetry axis. We have demonstrated this with our
Green's function theory in the RPA and shown that there can be
a sizeable difference in magnetic excitation energies at
finite fields. We show that the field dependence of the energy
is strongly temperature dependent and different for
field alignments along easy and hard directions.

It would be interesting to put our prediction of an anisotropy 
of the magnetic excitation energies in the paramagnetic phase to 
an experimental test. This may be possible in ultrathin magnetic films
using inelastic light scattering or ferromagnetic resonance measurements.

In fact, recent Brillouin light scattering experiments in Fe/GaAs(100) ultrathin films
($t_{Fe}=4.5$ \AA) showed anisotropy in observed frequencies above some
ordering temperature.\cite{preprint:2004} However, it is not clear from the results whether
the film is continuous or is instead a collection of superparamagnetic islands.

Our mean field arguments, for example, apply equally well to paramagnetic films above the Curie temperature and to
collections of superparamagnetic particles above the blocking temperature. The only difference might be that a
superparamagnetic "film" would be comprised of a distribution of particle sizes, each with a different local internal
magnetic field. Resonance in the superparamagnetic system would contain many different frequency contributions and
should therefore have a correspondingly broad linewidth. Resonance in the paramagnetic continuous film would
have a linewidth determined by higher moments in the spin correlation functions, and should therefore also have a
distinctive temperature dependence different from the superparamagnetic case.

\begin{acknowledgments}
R. L. Stamps acknowledges financial support from the Italian
Ministery for the Instruction, University and Research (MIUR), under
project FIRB no. RBNE017XSW. This work was performed in the framework of the joint
CNR-MIUR programme (legge 16/10/2000, Fondo FISR).
\end{acknowledgments}

\appendix

\section{Green's function formalism}
In order to calculate the equilibrium configuration of the monolayer
magnetization and the energies of the
magnetic excitations, we consider the following Green's functions
\begin{equation}
G_{ij}^{\alpha}=\langle\langle S_i^{\alpha}; S_j^-\rangle\rangle;
~~\alpha=+, -, z
\end{equation}
\noindent where $S_i^{\pm}=S_i^x \pm i S_i^y$ and
$x,y,z$ denote a reference frame where $z$ is the equilibrium
direction of the in-plane magnetization, forming an angle $\phi$
with the easy in-plane direction ${\it Z}$ (see Fig.~1).
The equations of motion for the time Fourier-transformed Green's functions
$G_{ij}^{\alpha}(\omega)$ are
\begin{equation}
\omega G_{ij}^{\alpha}(\omega)=\langle [S_i^{\alpha},S_j^-]\rangle
+\langle\langle [S_i^{\alpha},{\cal H}];S_j^-\rangle\rangle_{\omega}
\end{equation}
\noindent where $\langle A \rangle={1\over {\cal Z}} Tr(e^{-\beta{\cal H}} A)$
denotes a thermal average ($\beta={1\over {k_B T}}$ and 
${\cal Z}=Tr (e^{-\beta{\cal H}})$ is the partition function).
Taking into account that $S^{{\it X}}=S^x \cos \phi + S^z \sin\phi$
and $S^{{\it Z}}=-S^x \sin \phi + S^z \cos \phi$, one obtains
\begin{eqnarray}
&\omega& G_{ij}^{\pm}(\omega)=\langle [S_i^{\pm},S_j^-]\rangle
\cr &\mp& J \sum_k \langle\langle S_i^z S_k^{\pm}-S_k^zS_i^{\pm};S_j^-
\rangle\rangle_{\omega}
\cr &\pm& K_2 \cos^2 \phi \langle\langle
S_i^z S_i^{\pm}+S_i^{\pm}S_i^z; S_j^-
\rangle\rangle_{\omega}
%
%
\cr &\mp& {1\over 2} K_2 \sin^2 \phi \langle\langle
S_i^z S_i^- +S_i^- S_i^z+S_i^z S_i^+ +S_i^+ S_i^z ; S_j^-
\rangle\rangle_{\omega}
\cr &\pm& K_2 \sin\phi \cos\phi
\cr &&\langle\langle
2 S_i^z S_i^z -S_i^{\pm} S_i^{\pm} - {1\over 2}(S_i^+ S_i^-+S_i^- S_i^+) ; S_j^-
\rangle\rangle_{\omega}
\cr &\pm& g \mu_B H^{{\it Z}} ( \sin \phi ~G_{ij}^z(\omega)+\cos\phi ~G_{ij}^{\pm}(\omega))
\cr &\pm& g \mu_B H^{{\it X}} ( \sin \phi~ G_{ij}^{\pm}(\omega)-\cos\phi ~G_{ij}^z(\omega))
\end{eqnarray}

\begin{eqnarray}
&\omega& G_{ij}^{z}(\omega)=\langle [S_i^{z},S_j^-]\rangle
\cr &+& {1\over 2} J \sum_k \langle\langle S_i^- S_k^{+}-S_k^-S_i^{+};S_j^-
\rangle\rangle_{\omega}
\cr &-& {1\over 2} K_2 \sin^2 \phi \langle\langle
S_i^+ S_i^+ -S_i^- S_i^-; S_j^-
\rangle\rangle_{\omega}
\cr &+& {1\over 2} K_2 \sin\phi \cos\phi
\cr && \langle\langle
S_i^+ S_i^z +S_i^z S_i^+ - S_i^- S_i^z-S_i^z S_i^- ; S_j^-
\rangle\rangle_{\omega}
\cr &+& {1\over 2}g \mu_B H^{{\it Z}}  \sin \phi (G_{ij}^+(\omega)-G_{ij}^-(\omega))
\cr &-& {1\over 2}g \mu_B H^{{\it X}}  \cos \phi ( G_{ij}^+(\omega)-G_{ij}^-(\omega))
\end{eqnarray}

The higher order Green's functions on the right hand sides of the previous equations
have now to be decoupled to obtain a closed set of equations.
We start using the na\"ive RPA decoupling ($\alpha,\beta=+,-,z$)
for all the Green's functions ({\it i.e.}, both for $i \ne k$ and $i=k$;
kinematic consistency corrections will be taken into account later on)
\begin{equation}
\langle\langle
S_i^{\alpha} S_k^{\beta}; S_j^-
\rangle\rangle_{\omega} \approx \langle S_i^{\alpha}\rangle G_{kj}^{\beta}(\omega)
+\langle S_k^{\beta}\rangle G_{ij}^{\alpha}(\omega)
\end{equation}
Next, taking into account that the local axis $z$ is, by its own definition,
the equilibrium direction of the magnetization, we can safely put $\langle S^+ \rangle \approx 0$
and $\langle S^- \rangle \approx 0$ in the equations of motion of the
three Green's functions $G_{ij}^+(\omega)$, $G_{ij}^-(\omega)$ and $G_{ij}^z(\omega)$.

The equation of motion for $G_{ij}^z(\omega)$ provides the
equilibrium configuration. In fact in the previous approximations we obtain
\begin{equation}
\label{Gz}
\omega G_{ij}^z(\omega)\approx {1\over 2} \left\lbrack G_{ij}^+(\omega)-G_{ij}^-(\omega) \right\rbrack
\times \left\lbrack
K_2 \langle S_i^z \rangle \sin (2\phi)
+g \mu_B (H^Z \sin \phi - H^X \cos \phi)
\right\rbrack
\end{equation}
so that the equilibrium angle $\phi$ is obtained imposing the vanishing of the term in
braces on the r.h.s.
\begin{equation}
\label{equilibrium}
 2 K_2 \langle S_i^z \rangle \sin \phi \cos \phi
+g \mu_B (H^Z \sin \phi - H^X \cos \phi)=0
\end{equation}

For field applied along the easy axis ($H^{{\it Z}} \ne 0$ and $H^{{\it X}}=0$),
one obviously obtains $\phi=0$: {\it i.e.}, the magnetization is always
directed along the field (${\it Z}=z$) direction.
For field applied along the hard axis ($H^{{\it Z}}=0$ and $H^{{\it X}}\ne 0$),
the equilibrium angle is given by
\begin{eqnarray}
\phi &=&\sin^{-1} ({{g \mu_BH^{{\it X}}}\over {2 K_2 \langle S_i^z \rangle}}),~{\rm for}~H^X \le H^X_{c}
\cr
\phi &=& {{\pi}\over 2},~{\rm for}~H^X>H^X_{c}
\end{eqnarray}
so that a reorientation of the magnetization within the film plane
occurs at a critical value of the applied field
$H^{{\it X}}_{c}={{2 K_2 \langle S_i^z \rangle}\over {g\mu_B}}$
which turns out to be temperature dependent.
For $H^X>H^X_{c}$, the magnetization becomes parallel
to the external field $H^{{\it X}}$ applied along the hard direction.

The equations of motion for $G_{ij}^+(\omega)$ and $G_{ij}^-(\omega)$
provide the frequencies of the excitations with respect to the equilibrium
configuration. In fact for the space Fourier-transformed
Green's functions $G^{\pm}(\omega,{\bf k}_{\Vert})={1\over N}\sum_{ij}
G_{ij}^{\pm}(\omega)e^{ -i {\bf k}_{\Vert} \cdot {\bf r}_{ij} }$
(where ${\bf k}_{\Vert}=(k_x,k_z)$) we obtain
\begin{eqnarray}
&&\omega G^{\pm}(\omega,{\bf k}_{\Vert})\approx F^{\pm}({\bf k}_{\Vert})
\cr && \pm 4J\langle S_i^z \rangle ( 1-\gamma_{{\bf k}_{\Vert}} )
G^{\pm}(\omega,{\bf k}_{\Vert})
\cr && \pm 2 K_2 \langle S_i^z \rangle \cos^2 \phi G^{\pm}(\omega,{\bf k}_{\Vert})
\cr && \mp K_2 \langle S_i^z \rangle \sin^2 \phi \left\lbrack
G^+(\omega,{\bf k}_{\Vert})+G^-(\omega,{\bf k}_{\Vert})
\right\rbrack
\cr && \pm g \mu_B \left\lbrack
H^{{\it Z}} \cos \phi + H^{{\it X}} \sin \phi
\right\rbrack G^{\pm}(\omega,{\bf k}_{\Vert})
\end{eqnarray}
where $F^{\pm}({\bf k}_{\Vert})={1\over N}\sum_{ij}
e^{ -i {\bf k}_{\Vert} \cdot {\bf r}_{ij} }
\langle [S_i^{\pm},S_j^-] \rangle$ and 
$\gamma_{{\bf k}_{\Vert}}={1\over {z_{nn}}}\sum_{\delta} e^{-i {\bf k}_{\Vert}\cdot \delta }$
($\delta$ are the vectors joining a given lattice site to its $z_{nn}$ nearest neighbors).
For a square lattice with lattice constant $a$ one has $z_{nn}=4$ and the geometrical 
factor takes the form $\gamma_{{\bf k}_{\Vert}}={1\over 2}[\cos(k_x a)+\cos(k_z a)]$.

The previous equations can be rewritten in a more compact form
\begin{equation}
\label{matricial}
\left\lbrack
\begin{array}{l}
       \omega- A_{{\bf k}_{\Vert}} \;~~ B_{{\bf k}_{\Vert}} \;
        \cr
        -B_{{\bf k}_{\Vert}}~~ \; \omega+A_{{\bf k}_{\Vert}} \;
\end{array}
\right\rbrack
\left\lbrack
\begin{array}{l}
         G^+(\omega,{\bf k}_{\Vert}) \;
        \cr
        G^-(\omega,{\bf k}_{\Vert}) \;
\end{array}
\right\rbrack =
\left\lbrack
\begin{array}{l}
         F^+({\bf k}_{\Vert}) \;
        \cr
        F^-({\bf k}_{\Vert}) \;
\end{array}
\right\rbrack
\end{equation}
where the quantities $A_{{\bf k}_{\Vert}}$ and $B_{{\bf k}_{\Vert}}$ take the
following expressions
\begin{eqnarray}
\label{AkBk}
A_{{\bf k}_{\Vert}}&=&
4J\langle S_i^z \rangle(1-\gamma_{{\bf k}_{\Vert}})+2K_2 \langle S_i^z\rangle \cos^2 \phi
- K_2 \langle S_i^z \rangle \sin^2 \phi +g\mu_B
(H^{{\it Z}}\cos\phi+H^{{\it X}}\sin\phi)
\cr
B_{{\bf k}_{\Vert}}&=&
K_2 \langle S_i^z \rangle \sin^2 \phi
\end{eqnarray}
The energies of the magnetic excitations with respect to the
equilibrium state are then given by
\begin{equation}
\label{frequencies}
E_{{\bf k}_{\Vert}}^2= (\hbar \omega_{{\bf k}_{\Vert}})^2=A_{{\bf k}_{\Vert}}^2 -B_{{\bf k}_{\Vert}}^2
\end{equation}
It is worthwhile noticing that in the zero temperature limit
one has $\langle S_i^z \rangle  \to S$ so that the free
spin wave frequency is correctly recovered.
In particular, in the $T=0$ limit we are able to recover the well-known feature
of a zero frequency mode\cite{morrish} for field applied along the hard axis and
equal to the critical reorientation field, in contrast with the results obtained 
by the Green's function approach in Ref.~\onlinecite{nolting}. 
It is worth noticing that other Green's function 
approaches\cite{froebrich,froebrich2,froebrich3} 
were not able to recover such a feature, either.

So far, in our calculations, the problem of the kinematic consistency of
the magnetic excitations was neglected since the na\"ive
RPA decoupling was performed also for the Green's functions coming
from the anisotropy term in the Hamiltonian.
In the case of field applied along the easy axis ($H^{{\it Z}}\ne 0$, $H^{{\it X}}=0$),
it is well known that a more correct treatment of the anisotropy term
is obtained by using the Anderson-Callen decoupling to decouple the equal site ($i=k$)
Green's functions\cite{callen}
\begin{equation}
\langle\langle S_i^{\pm} S_i^z + S_i^z S_i^{\pm}; S_j^-
\rangle\rangle_{\omega} \approx 2 \langle S_i^z \rangle G_{ij}^{\pm}(\omega)
\times \left\lbrace
1-{1\over {2S^2}} \left\lbrack
S(S+1) - \langle S_i^z S_i^z \rangle
\right\rbrack
\right\rbrace
\end{equation}
In this way, for $S=1/2$, the uniaxial anisotropy does not contribute
to the frequency of the excitations, as expected, since for $S=1/2$ the
anisotropy term in the Hamiltonian is a constant.
In the case of field applied along the hard axis ($H^{{\it Z}}=0$, $H^{{\it X}}\ne 0$)
or in the general case ($H^{{\it Z}}\ne 0$ and $H^{{\it X}}\ne 0$), we assume
the same prescription for kinematic consistency to be valid as in the easy case,
{\it i.e.} we replace
\begin{equation}
\label{kinem_anyspin}
K_2 \langle S_i^z \rangle \to K_2 \langle S_i^z \rangle
\left\lbrace
1-{1\over {2S^2}} \left\lbrack
S(S+1) - \langle S_i^z S_i^z \rangle
\right\rbrack
\right\rbrace
\end{equation}
both in the equation (\ref{equilibrium}) providing the equilibrium condition and in the
equations (\ref{AkBk}) and (\ref{frequencies}) giving the energies of the magnetic excitations.
In the case $S=1$ considered in this paper, the prescription for kinematic
consistency reads simply
\begin{equation}
\label{kinemspin1}
2 K_2 \langle S_i^z \rangle \to K_2 \langle S_i^z \rangle \langle S_i^z S_i^z \rangle
\end{equation}
In order to determine the average $\langle S_i^z S_i^z \rangle$, we follow the method
exposed by Callen\cite{callen} in his paper. In addition to the Green's functions
$G_{ij}^{\pm}=\langle\langle S_i^{\pm}; S_j^- \rangle\rangle$, we consider
the Green's functions
${\cal G}_{ij,a}^{\pm}=\langle\langle e^{a S_i^z} S_i^{\pm}; S_j^- \rangle\rangle$
where $a$ is a parameter.  The ${\cal G}$'s are found to satisfy the equations of motion
\begin{equation}
\label{matriciale}
\left\lbrack
\begin{array}{l}
       \omega- A({\bf k}_{\Vert}) \;~~ B({\bf k}_{\Vert}) \;
        \cr
        -B({\bf k}_{\Vert})~~ \; \omega+A({\bf k}_{\Vert}) \;
\end{array}
\right\rbrack
\left\lbrack
\begin{array}{l}
         {\cal G}_a^+(\omega,{\bf k}_{\Vert}) \;
        \cr
        {\cal G}_a^-(\omega,{\bf k}_{\Vert}) \;
\end{array}
\right\rbrack =
\left\lbrack
\begin{array}{l}
         {\cal F}_a^+({\bf k}_{\Vert}) \;
        \cr
        {\cal F}_a^-({\bf k}_{\Vert}) \;
\end{array}
\right\rbrack
\end{equation}
which differ from the equation of motion for the $G$'s only for the terms
${\cal F}_a^{\pm}$ on the r.h.s., defined as 
${\cal F}_a^{\pm}({\bf k}_{\Vert})={1\over N}\sum_{ij}
e^{ -i {\bf k}_{\Vert} \cdot {\bf r}_{ij} }
\langle [S_i^{\pm},e^{aS^z_j}S_j^-] \rangle$.
Thus from the spectral theorem
\begin{equation}
\langle B A \rangle = {{i}\over {2\pi}} \lim_{\epsilon \to 0}
\int_{-\infty}^{+\infty} {{d\omega}\over {e^{\beta\omega}-1}}
\left\lbrack
\langle \langle A;B \rangle\rangle (\omega+i\epsilon,{\bf k}_{\Vert})
-\langle \langle A;B \rangle\rangle (\omega-i\epsilon,{\bf k}_{\Vert})
\right\rbrack
\end{equation}
applied to $G^+$ and ${\cal G}_a^+$
one can obtain, respectively, the thermal averages
\begin{eqnarray}
\langle S_i^- S_i^+\rangle&=&F^+({\bf k}_{\Vert})~ \Phi(T)
\cr
\langle e^{a S_i^z}S_i^- S_i^+ \rangle&=&{\cal F}_a^+({\bf k}_{\Vert}) ~\Phi(T)
\end{eqnarray}
where
\begin{equation}
\Phi(T)={1\over {2N}}\sum_{{\bf k}_{\Vert}} \left\lbrack
{{A_{{\bf k}_{\Vert}}}\over {E_{{\bf k}_{\Vert}}}}
\coth ({1\over 2} \beta E_{{\bf k}_{\Vert}}) -1
\right\rbrack
\end{equation}
For the two-spin correlation function we obtain
\begin{equation}
\langle S_i^- S_i^+ \rangle= 2 \langle S_i^z \rangle \Phi(T)
\end{equation}
Following Callen,\cite{callen} it is now possible to find a second-order
differential equation for ${\cal F}_a^+({\bf k}_{\Vert})$ as a function of the parameter $a$
that, when solved with opportune boundary conditions,
allows to obtain the magnetization as a function of $\Phi(T)$
\begin{equation}
\langle S_i^z \rangle=
{{[S-\Phi(T)][1+\Phi(T)]^{2S+1}+[S+1+\Phi(T)][\Phi(T)]^{2S+1}
}\over {[1+\Phi(T)]^{2S+1}-[\Phi(T)]^{2S+1}}}
\end{equation}
and finally
\begin{equation}
\langle S_i^z S_i^z \rangle = S(S+1) - \langle S_i^- S_i^+ \rangle - \langle S_i^z \rangle
=S(S+1)-(1+2 \Phi(T))\langle S_i^z \rangle
\end{equation}

\section{Free spin wave theory}
We think it useful to briefly present here the results of non interacting 
spin wave theory, both for the equilibrium condition and the energy of the 
magnetic excitations in an anisotropic monolayer described by the microscopic 
Hamiltonian (\ref{hamiltonian}), since in the $T \to 0$ limit the spin wave 
results are expected to be exactly reproduced by Green's function theory.
Denoting by $x,y,z$ the local reference frame where $z$ is the equilibrium
direction of the in-plane magnetization, forming an angle $\phi$
with the easy in-plane direction ${\it Z}$ (see Fig.~1), 
the local spin components are expressed in terms of boson operators
via the Holstein-Primakoff transformation\cite{spin-wave-theory}
\begin{equation}
\label{hp}
S^+_j=\sqrt{2S}\sqrt{1-{{a^+_j a_j}\over {2S}}} a_j,~~
S^-_j=\sqrt{2S} a^+_j\sqrt{1-{{a^+_j a_j}\over {2S}}},~~ 
S^z_j=S-a^+_j a_j
\end{equation}
while the spin components in the crystallographic reference frame 
are obtained from the local ones through the equations 
$S^{{\it X}}=S^x \cos \phi + S^z \sin\phi$
and $S^{{\it Z}}=-S^x \sin \phi + S^z \cos \phi$.
In the following, for the sake of simplicity,  
we neglect both dipolar interactions and kinematic consistency corrections 
to the anisotropy term in Eq. (\ref{hamiltonian}).
Substituting in (\ref{hamiltonian}), expanding the square roots in Eq.~(\ref{hp}) 
(such an approximation is valid for low values of the occupation number $n_j=a^+_j a_j$, 
that is for low temperatures), and keeping only up to 
quadratic terms in the boson operators one obtains, after a space Fourier 
transformation $a_j={1\over {\sqrt{N}}}\sum_{{\bf k}_{\Vert}} a_{{\bf k}_{\Vert}}
e^{ -i {\bf k}_{\Vert} \cdot {\bf r}_{ij} }$
\begin{eqnarray}
{\cal H}&\approx&{\cal H}_1+{\cal H}_2= 
\sqrt{{S\over 2}}\sum_{{\bf k}_{\Vert}}(a_{{\bf k}_{\Vert}}+a^+_{{\bf k}_{\Vert}})
\left\lbrack
2 K_2 S \sin \phi \cos\phi + g \mu_B (H^Z \sin \phi - H^X \cos \phi)
\right\rbrack
\cr
&+&\sum_{{\bf k}_{\Vert}} \left\lbrack
A_{{\bf k}_{\Vert}} a_{{\bf k}_{\Vert}}^+ a_{{\bf k}_{\Vert}} 
+ {1\over 2} B_{{\bf k}_{\Vert}}
(a_{{\bf k}_{\Vert}} a_{-{{\bf k}_{\Vert}}}+a^+_{{\bf k}_{\Vert}} a^+_{-{{\bf k}_{\Vert}}})
\right\rbrack
\end{eqnarray}
where 
\begin{eqnarray}
A_{{\bf k}_{\Vert}}&=&4JS (1-\gamma_{{\bf k}_{\Vert}})+2 K_2 S \cos^2 \phi - K_2 S \sin^2 \phi
+g\mu_B (H^X \sin \phi+H^Z\cos\phi)
\cr
B_{{\bf k}_{\Vert}}&=& -K_2 S \sin^2 \phi.
\end{eqnarray}
\noindent The equilibrium condition is obtained by 
imposing the vanishing of the one-boson Hamiltonian ${\cal H}_1$
\begin{equation}
\label{equilsw}
2K_2 S \sin \phi \cos \phi + g \mu_B (H^Z \sin \phi -H^X \cos \phi)=0
\end{equation}
It is readily seen that the spin wave theory result  
Eq. (\ref{equilsw}) is reproduced by Green's function theory 
in the limit $T \to 0$, see Eq. (\ref{equilibrium}).  

The two-boson Hamiltonian ${\cal H}_2$ can be readily diagonalized\cite{Yafet,Bruno,noi}
and the energy of the non interacting spin wave excitations turns out to be
\begin{equation}
E_{{\bf k}_{\Vert}}=\sqrt{A_{{\bf k}_{\Vert}}^2 -B_{{\bf k}_{\Vert}}^2}
\end{equation}
Taking into account both dipolar interactions\cite{Yafet,Bruno,noi} and 
kinematic consistency corrections,\cite{balucani}
for field $H^X$ applied along the hard in-plane direction
one finally obtains the free spin wave energy explicitly 
reported in Eq. (\ref{E0hard}).

\section{Contribution of dipolar interactions}
The inclusion of magnetic dipole-dipole interactions ({\it i.e.}, the
last term in Eq.~(\ref{hamiltonian})) in the equations of motion for the
Green's functions leads to a considerable amount of additional calculations.
As they are tedious but straightforward, hereafter we only quote the final result
after RPA decoupling.
First we consider the commutator of ${\cal H}_{dip}$ with $S_i^z$
in order to investigate the contribution of the dipolar interaction to the equilibrium
condition. We have ($w={{g^2 \mu_B^2}\over {a^3}}$)
\begin{eqnarray}
&&\langle\langle [S_i^z,{\cal H}_{dip}];S_j^- \rangle\rangle_{\omega}
\approx -{3\over 2} w \sum_{k\ne l}
\Big\{ \langle S_l^z \rangle \delta_{ik}+\langle S_k^z \rangle \delta_{il}
\Big\}
\cr
&& \Big\{ \left\lbrack
(d_{kl}^{XX}-d_{kl}^{ZZ})\sin\phi \cos\phi +d_{kl}^{XZ}(\cos^2\phi-\sin^2\phi)\right\rbrack
\times {{G_{ij}^+(\omega)-G_{ij}^-(\omega)}\over 2}
\cr &+&  \left\lbrack d_{kl}^{XY}\sin\phi  +d_{kl}^{YZ}\cos\phi  \right\rbrack 
\times {{G_{ij}^+(\omega)+G_{ij}^-(\omega)}\over {2i}}
\Big\}
\end{eqnarray}
where the dipolar sums are defined as ($\alpha,\beta=X,Y,Z$)
\begin{equation}
D^{\alpha\beta}({\bf k}_{\Vert})=\sum_{i\ne j} d_{ij}^{\alpha\beta} e^{i {\bf k}_{\Vert}\cdot {\bf r}_{ij}}
=\sum_{i\ne j} {{a^3}\over {r_{ij}^3}}~{{r_{ij}^{\alpha} r_{ij}^{\beta}}\over {r_{ij}^2}}
e^{i {\bf k}_{\Vert}\cdot {\bf r}_{ij}}
\end{equation}
For a monolayer with the crystallographic $Y$ axis normal to the film plane
one has $r_{kl}^Y=0$ so that $D^{XY}=D^{YZ}=0$; these relations are approximately
satisfied even in an ultrathin film. Moreover, in a lattice with a center of
inversion symmetry one has $D^{XX}=D^{ZZ}$ and $D^{XZ}=0$.
Thus, as expected, it is proved that the dipolar interaction
does not contribute to the equilibrium condition of an in-plane magnetized
ultrathin film.

Next, we consider the commutator of ${\cal H}_{dip}$ with $S_i^{\pm}$
in order to obtain the contribution of the dipolar interaction to the magnon energies.
In the same approximations as before we obtain, after a Fourier transformation
\begin{eqnarray}
A_{{\bf k}_{\Vert}}^{dip}&=&{1\over 2} w \langle S_i^z \rangle
\sum_{i\ne j} {{a^3}\over {r_{ij}^3}} e^{i {\bf k}_{\Vert}\cdot {\bf r}_{ij}}
\times \Bigg\{1+
\left\lbrack 1- 3\Big({{r_{ij}^X\cos\phi-r_{ij}^Z\sin\phi}\over {r_{ij}}}\Big)^2
\right\rbrack
\cr &-&
2\left\lbrack 1- 3\Big({{r_{ij}^X\sin\phi+r_{ij}^Z\cos\phi}\over {r_{ij}}}\Big)^2
\right\rbrack
\Bigg\}
\cr
B_{{\bf k}_{\Vert}}^{dip}&=&{1\over 2} w \langle S_i^z \rangle
\sum_{i\ne j} {{a^3}\over {r_{ij}^3}} e^{i {\bf k}_{\Vert}\cdot {\bf r}_{ij}}\times
\Bigg\{1-
\left\lbrack 1- 3\Big({{r_{ij}^X\cos\phi-r_{ij}^Z\sin\phi}\over {r_{ij}}}\Big)^2
\right\rbrack
\Bigg\}
\end{eqnarray}
\noindent In order to simplify the numerical calculations, we neglect the dependence
of $A_{{\bf k}_{\Vert}}^{dip}$ and $B_{{\bf k}_{\Vert}}^{dip}$ on ${\bf k}_{\Vert}$.
Taking into account that for a two-dimensional lattice with a center 
of inversion symmetry one has 
\begin{equation}
\sum_{i\ne j} {{a^3}\over {r_{ij}^3}}~\Big( { {r_{ij}^{X}}\over {r_{ij}} } \Big)^2
=\sum_{i\ne j} {{a^3}\over {r_{ij}^3}}~\Big({{r_{ij}^{Z}}\over {r_{ij}}}\Big)^2
= -{1\over 2} \sum_{i\ne j} {{a^3}\over {r_{ij}^3}}=-{4\over 3} \pi f_w
\end{equation}
with $f_w=1.078$ for the s.q. lattice,\cite{Yafet,Bruno,noi} 
we thus obtain the approximate expressions
(\ref{Ak}) and (\ref{Bk}) reported in Section II.
Clearly, such an approximation on the dipolar part of the Hamiltonian
is allowed as long as the strength of the dipolar interaction $w$ is smaller than
the intensity of the uniaxial anisotropy and/or the Zeeman term.


\begin{thebibliography}{99}

\bibitem{back:1994}
C. H. Back, A. Kashuba, D. Pescia, Phys. Low-Dim. Struct. {\bf 2}, 9 (1994).

\bibitem{jensen:2003}
P. J. Jensen, S. Knappmann, W. Wulfhekel, H. P. Oepen, Phys. Rev. B {\bf 67}, 184417 (2003).

\bibitem{demokritov}
S. O. Demokritov, N. M. Kreines, V. I. Kudinov, S. V. Petrov, 
Sov. Phys. JETP {\bf 68}, 1277 (1989).

\bibitem{dutcher}
J. R. Dutcher, B. Heinrich, J. F. Cochran, D. A. Steigerwald, W. F. Egelhoff, Jr.,
J. Appl. Phys. {\bf 63}, 3464 (1988).

\bibitem{spin-wave-theory}
T. Holstein and H. Primakoff, Phys. Rev. {\bf 58}, 1098 (1940).

\bibitem{Yafet}
Y. Yafet, J. Kwo, E. M. Gyorgy, Phys. Rev. B {\bf 33}, 6519 (1986).

\bibitem{Bruno}
P. Bruno, Phys. Rev. B {\bf 43}, 6015 (1991). 

\bibitem{noi}
P. Politi, A. Rettori, M. G. Pini, J. Magn. Magn. Mater. {\bf 113}, 83 (1992).

\bibitem{Pescia}
M. G. Pini, A. Rettori, D. Pescia, N. Majlis, S. Selzer, in 
{\it Microscopic Aspects of Nonlinearity in Condensed Matter}, 
edited by A. R. Bishop, V. L. Pokrovsky and V. Tognetti (Plenum, New 
York, 1991); M. G. Pini, A. Rettori, D. Pescia, N. Majlis, S. Selzer, 
Phys. Rev. B {\bf 45}, 5037 (1992).

\bibitem{note1}
In this geometry the thermal average in the field direction is
$\langle S^{X}\rangle$. This average does not appear in the
effective field in the case of a continuous thin film,
but could appear if additional in plane shape factors existed.

\bibitem{usadel}
A. Hucht and K. D. Usadel, Phil. Mag. B {\bf 80}, 275 (2000).
K. D. Usadel and A. Hucht, Phys. Rev. B {\bf 66}, 024419 (2002).

\bibitem{devlin}
J. F. Devlin, Phys. Rev. B {\bf 4}, 136 (1971).


\bibitem{zubarev}
D. N. Zubarev, Usp. Fiz. Nauk. {\bf 71}, 71 (1960) [Sov. Phys. Usp. {\bf 3}, 320 (1960)].

\bibitem{froebrich}
P. Fr\"obrich, P. J. Jensen, P. J. Kuntz, Eur. Phys. J. B {\bf 13}, 477
(2000).
\bibitem{froebrich2}
P. Fr\"obrich, P. J. Jensen, P. J. Kuntz, A. Ecker, Eur. Phys. J. B
{\bf 18}, 579 (2000)
\bibitem{froebrich3}
P. Fr\"obrich, P. J. Kuntz, J. Phys.: Condens. Matter {\bf 16}, 3453
(2004).
\bibitem{froebrich_annals}
P. Fr\"obrich, P. J. Kuntz, S. Saber, Ann Phys. (Leipzig) {\bf 11}, 387 (2002).

\bibitem{nolting}
S. Schwieger, J. Kienert, W. Nolting, Phys. Rev. B {\bf 71}, 024428 (2005).

\bibitem{callen}
H. B. Callen, Phys. Rev. {\bf 130}, 890 (1963);
F. B. Anderson, H. B. Callen, Phys. Rev. {\bf 136}, A1068 (1964).


\bibitem{henelius}
P. Henelius, P. Fr\"obrich, P. J. Kuntz, C. Timm, P. J. Jensen, 
Phys. Rev. B {\bf 66}, 094407 (2002).

\bibitem{note_nolting_1}
Compare Eq. (18) of Ref. \onlinecite{nolting}
with Eq. (\ref{eq_main}) of the present paper.

\bibitem{note_nolting_2}
Compare Eq. (19) of Ref. \onlinecite{nolting} with Eqs. (\ref{magnon}), (\ref{Ak}), 
(\ref{Bk}) of the present paper.

\bibitem{note_nolting_3}
In fact, from Eq. (19) of Ref. \onlinecite{nolting} one has 
that for zero wavevector the magnon energy reduces to the effective field 
$B$ so that, substituting the condition for the magnetization angle 
obtained from Eq. (18) of Ref. \onlinecite{nolting}, one finds 
that, in the $T \to 0$ limit (when the thermal averages 
$\langle S_{z\prime}\rangle$ and $\langle S_{z\prime}^2\rangle$ 
tend to constant values), the magnon energy does not vanish 
at the reorientation field. 

\bibitem{morrish}
A. H. Morrish, {\it The Physical Principles of Magnetism}, (Wiley, New York, 1965),
p. 555.



\bibitem{junger}
I. Junger, D. Ihle, J. Richter, A. Kl\"umper, Phys. Rev. B {\bf 70}, 104419 (2004).

\bibitem{preprint:2004}
G. Carlotti {\it et al.}, private communication (2004).

\bibitem{balucani}
U. Balucani, V. Tognetti, M. G. Pini, J. Phys. C.: Solid State Phys. 
{\bf 12}, 5513 (1979).

\end{thebibliography}
\end{document}